\documentclass[10pt]{article}
\usepackage{multirow}
\usepackage[dvips]{graphicx}

\usepackage[psamsfonts]{amssymb}
\usepackage{amsxtra}
\usepackage{threeparttable}
\usepackage{amsmath}
\usepackage{amssymb}

\usepackage{graphicx}

\usepackage{cite}

\usepackage{color}

\usepackage{setspace}
\usepackage{float}

\usepackage{url}

\usepackage[labelfont=bf,labelsep=period,justification=raggedright]{caption}

\makeatletter
\renewcommand{\@biblabel}[1]{\quad#1.}
\makeatother

\date{April 18, 2013}

\begin{document}

\begin{flushleft}
{\Large
\textbf{Multi-command Tactile Brain Computer Interface:\\ A Feasibility Study}
}

Hiromu Mori$^1$, 
Yoshihiro Matsumoto$^1$, 
Victor Kryssanov$^2$, 
Eric Cooper$^2$, 
Hitoshi Ogawa$^2$, \\
Shoji Makino$^1$, 
Zbigniew R. Struzik$^{3,4}$,
and Tomasz M. Rutkowski$^{1,4,}$\footnote{The corresponding author.}$^,$\footnote{The final publication is available at link.springer.com}
\\
\bf{1} Life Science Center of TARA, University of Tsukuba, Tsukuba, Japan\\
\bf{2} Ritsumeikan University, Shiga, Japan\\
\bf{3} University of Tokyo, Tokyo, Japan\\
\bf{4} RIKEN Brain Science Institute, Wako-shi, Japan\\
$\ast$ E-mail: tomek@tara.tsukuba.ac.jp
\end{flushleft}

\section*{Abstract}

The study presented explores the extent to which tactile stimuli delivered to the ten digits of a BCI-naive subject can serve as a platform for a brain computer interface (BCI) that could be used in an interactive application such as robotic vehicle operation. The ten fingertips are used to evoke somatosensory brain responses, thus defining a tactile brain computer interface (tBCI). Experimental results on subjects performing online (real-time) tBCI, using stimuli with a moderately fast inter-stimulus-interval (ISI), provide a validation of the tBCI prototype, while the feasibility of the concept is illuminated through information-transfer rates obtained through the case study.

\noindent{\bf Keywords:} tactile BCI, P300, robotic vehicle interface

\section{Introduction}

Contemporary BCIs are typically based on mental visual and motor imagery prototypes, which require extensive user training and non-impaired vision of the users~\cite{bciBOOKwolpaw}. Recently alternative solutions have been proposed to make use of spatial auditory~\cite{aBCIbirbaumer2010,bciSPATIALaudio2010} or tactile (somatosensory) modalities~\cite{sssrBCI2006,HBCIscis2012hiromuANDtomek,JNEtactileBCI2012} to enhance brain-computer interface comfort and increase the information transfer rate (ITR) achieved by users. The concept proposed in this paper uses the brain somatosensory (tactile) channel to allow targeting of the tactile sensory domain for the operation of robotic equipment such as personal vehicles, life support systems, etc. The rationale behind the use of the tactile channel is that it is normally far less loaded than visual or even auditory channels in such applications.  

One of the first reports~\cite{sssrBCI2006} of the successful employment of steady-state somatosensory responses 
to create a BCI targeted a low frequency vibrotactile stimulus in the range of $20-31$~Hz to evoke the subjects' attentional modulation, which was then used to define interfacing commands. A recent report~\cite{JNEtactileBCI2012} proposed using a Braille stimulator with $100$~ms static push stimuli delivered to each of six fingers to evoke a somatosensory response potential (SEP) related P300. The P300 response is a positive electroencephalogram event-related potential (ERP) deflection starting at around $300$~ms and lasting for $200-300$~ms after an expected stimulus in a random series of distractors (the so-called oddball EEG experimental paradigm)~\cite{book:eeg}. P300 responses are commonly used in BCI approaches and are considered to be the most reliable ERPs~\cite{bci2000book,bciBOOKwolpaw} with untrained subjects. The results indicated that the experiments achieved information transfer rates of $7.8$~bit/min on average and $27$~bit/min for the best subject. 

This paper proposes a novel tactile brain-computer interface based on P300 responses evoked by tactile stimuli delivered via vibrotactile exciters attached to the ten fingertips of the subject's hands.

The rest of the paper is organized as follows. The next section introduces the materials and methods used in the study and also outlines the experiments conducted. The results obtained in psychophysical and electroencephalogram experiments with eleven BCI-naive subjects are then discussed. Finally, conclusions are formulated and directions for future research are outlined. 

\section{Materials and Methods}

Eleven paid BCI-naive subjects (ten males and one female) participated in the experiments. The subjects' mean age was $21.82$, with a standard deviation of $0.87$. All the experiments were performed at the Life Science Center of TARA, University of Tsukuba, Japan. 
The psychophysical and online (real-time) EEG tBCI prototype experiments were conducted in accordance with the \emph{WMA Declaration of Helsinki - Ethical Principles for Medical Research Involving Human Subjects}. 

\subsection{Tactile Stimuli}

The tactile stimuli were delivered as sinusoidal waves generated by a portable computer with \textsf{MAX/MSP} software~\cite{maxMSP}. The stimuli were generated via ten channel outputs (one for each fingertip of the subject) of an external \emph{digital-to-analog} signal converter \textsf{MOTU UltraLite-mk3 Hybrid} coupled with three \textsc{YAMAHA P4050} power amplifiers (four acoustic frequency channels each). 

\begin{figure}[H]
	\begin{center}
	\includegraphics[width=\linewidth, clip]{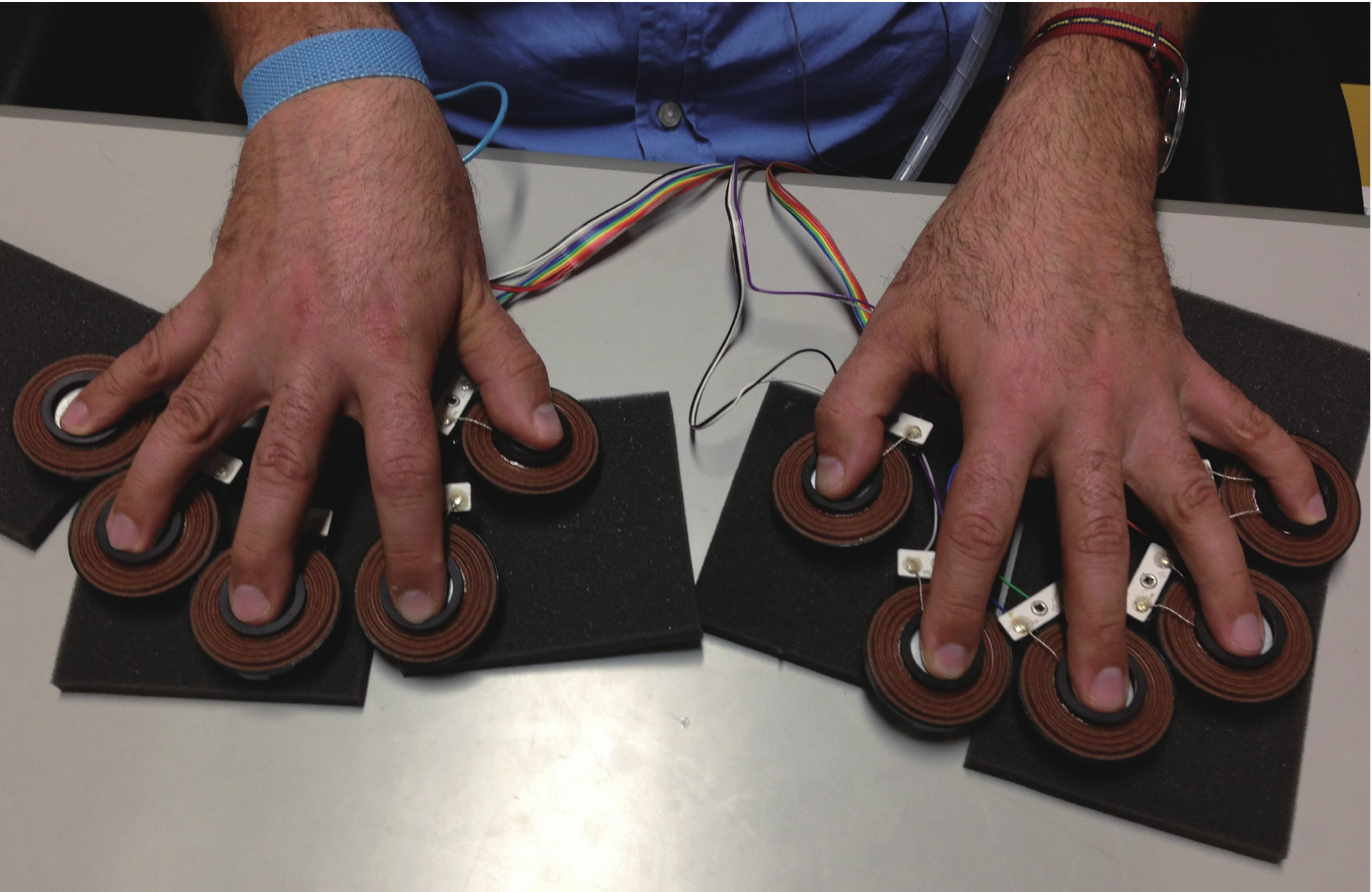}
	\end{center}
	\caption{The experimental set-up. Tactile stimuli are delivered to the fingertips by ten \textsf{HIAX25C10-8} exciters. The set-up was used in both the psychophysical and the EEG experiments.}
	\label{fig:10fingers}
\end{figure}

The stimuli were delivered to the subjects' fingertips via the tactile exciters \textsf{HiWave HIAX25C10-8/HS} working in the range of $100-20,000$~Hz, as depicted in Figure~\ref{fig:10fingers}. Each exciter in the experiments was set to emit a sinusoidal wave at $200$~Hz to match the exciter's resonance frequency and to stimulate the \emph{Pacini endings} (fast-adapting type II afferent type tactile sensory hand innervation receptors)~\cite{natureHAPTIC2009}.

\begin{figure}[H]
	\begin{center}
	\includegraphics[width=\linewidth, clip]{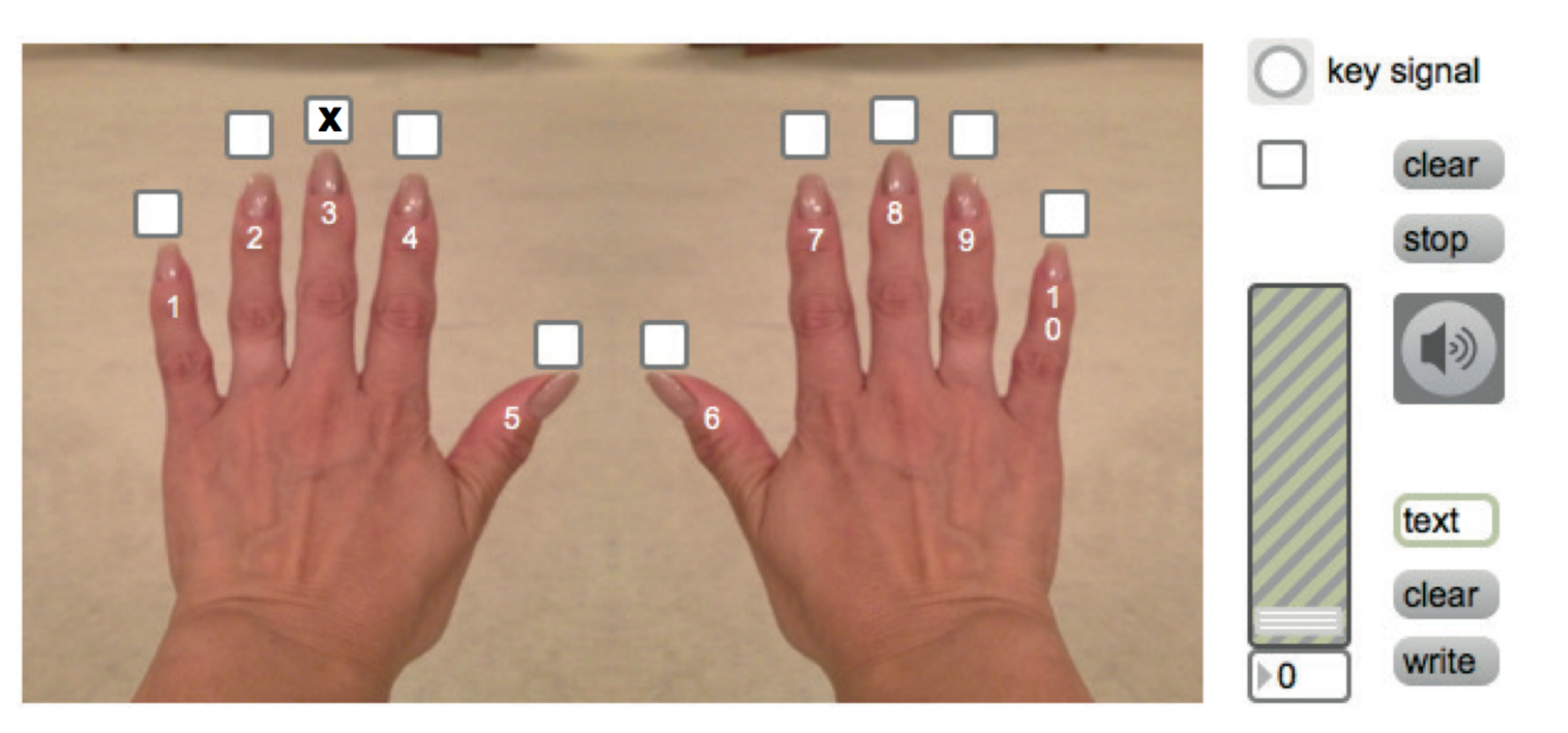}
	\end{center}
	\caption{The visual instruction screen presented to the subjects during the psychophysical experiment.  Each fingertip was assigned a number encoding a command in the interactive application. The controls on the right side were used by the subject to adjust the stimulus intensity (with a fader), and also by the experimenter to start the experiment (the button with the speaker launches the \emph{digital-to-analog} signal converter \textsf{MOTU UltraLite-mk3 Hybrid}) and to save/clear the results.}
	\label{fig:10fingersINTERFACE}
\end{figure}

The subjects placed their fingertips on the exciters (see Figure~\ref{fig:10fingers}) and attended only to the instructed locations (with a button-press response in the psychophysical experiments, or a mental count of the targets in the EEG experiments). The training instructions were presented visually by means of the \textsf{MAX/MSP} program, as depicted in Figure~\ref{fig:10fingersINTERFACE}.

\subsection{Psychophysical Experiment Protocol}

The psychophysical experiment was conducted to investigate the stimulus carrier frequency influence on the subjects' response time and accuracy.
The behavioral responses were collected using a small numeric keypad and the  \textsf{MAX/MSP} program. Each subject was instructed which stimulus to attend to by a cross above the target fingertip ($TARGET$), shown by the program (Figure~\ref{fig:10fingersINTERFACE}). Then the subject pressed the response button with the dominant foot. In the experiment, each subject was presented with $50~TARGETS$ and $450~non-TARGETS$ as stimuli. 

Each trial was composed of a randomized order of $100$~ms tactile bursts delivered to each fingertip separately with an inter-stimulus-interval (ISI) of $600$~ms. Every random sequence thus contained a single $TARGET$ and nine $non-TARGETS$. A single session included five trials for each $TARGET$ fingertip (resulting in $50~TARGETS$ and $450~non-TARGETS$). The choice of the relatively long ISI is justified by slow behavioral responses in comparison to EEG evoked potentials~\cite{book:eeg}, as described in the next section.

The response time delays were registered with the same \textsf{MAX/MSP} program, also used for the stimulus generation and instruction presentation.

\subsection{EEG tBCI Experiment}
 
\begin{figure}[H]
	\begin{center}
	\includegraphics[width=0.85\linewidth]{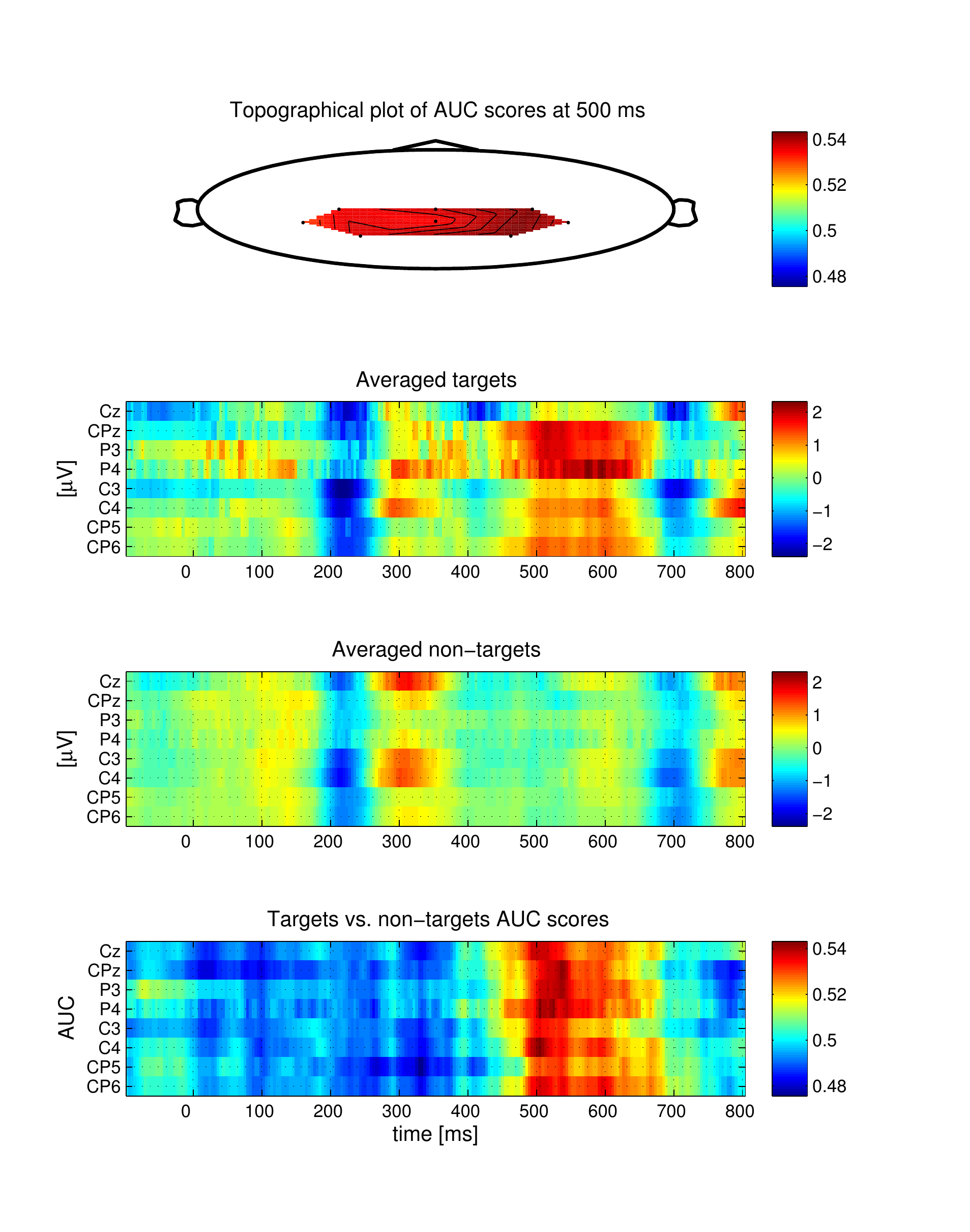}
	\end{center}
	\vspace{-1cm}
	\caption{Grand mean averaged results of the fingertip stimulation EEG experiment for all $11$ subjects. The top panel presents the head topographic plot of the $TARGET$ versus $non-TARGET$ area under the curve (AUC), a measure commonly used in machine learning intra-class discriminative analysis. ($AUC > 0.5$ is usually assumed to be confirmation of feature separability~\cite{book:pattRECO}). The top panel presents the largest difference as obtained from the data displayed in the bottom panel. The topographic plot also depicts the electrode positions. The fact that all the electrodes received similar AUC values (red) supports the initial electrode placement. The second panel from the top presents averaged SEP responses to the $TARGET$ stimuli (note the clear P300 response in the range of $450-700$~ms). The third panel presents averaged SEP responses to the $non-TARGET$ stimuli (no P300 observed). Finally, the bottom panel presents the AUC of $TARGET$ versus $non-TARGET$ responses (again, P300 could easily be identified).}\label{fig:feall}
\end{figure}

The exciters were attached to the fingertips in the same manner (see Figure~\ref{fig:10fingers}). During the experiment, EEG signals were captured with a portable wireless EEG amplifier system \textsf{g.MOBllab+} and \textsf{g.SAHARA} by \textsf{g.tec}, using eight dry electrodes. The electrodes were attached to the head locations: \textsf{Cz, CPz, P3, P4, C3, C4, CP5,} and \textsf{CP6}, as in the $10/10$ extended international system~\cite{Jurcak20071600} (see the topographic plot in the top panel of Figure~\ref{fig:feall}). The ground and reference electrodes were attached behind the left and right ears respectively. In order to limit electromagnetic interference, the subjects' hands were additionally grounded with armbands connected to the amplifier ground. No electromagnetic interference was observed from the exciters. Details of the EEG experimental protocol are summarized in Table~\ref{tb:p4}.

The recorded EEG signals were processed by a \textsf{BCI2000}-based application~\cite{bci2000book}, using a stepwise linear discriminant analysis (SWLDA) classifier~\cite{p300classifier} with features drawn from the $0-700$~ms ERP interval. The sampling rate was set at $256$~Hz, the high pass filter at $0.1$~Hz, and the low pass filter at $40$~Hz. The ISI was $400$~ms, and each stimulus duration was $100$~ms. The subjects were instructed to spell out the number sequences (corresponding to the interactive robotic application commands shown in Table~\ref{tab:commands}) communicated by the exciters in each session. Each $TARGET$ was presented five times in a single command trial, and the averages of the five ERPs were later used for the classification. Each subject performed three experimental sessions (randomized $50~TARGETS$ and $450~non-TARGETS$ each), which were later averaged as discussed in Section~\ref{sec:results}.

\section{Results}\label{sec:results}

This section discusses the results obtained in the psychophysical experiment and in the EEG experiment. 

\subsection{Psychophysical Experiment Results}

\begin{figure}[H]
	\begin{center}
	\includegraphics[width=\linewidth, clip]{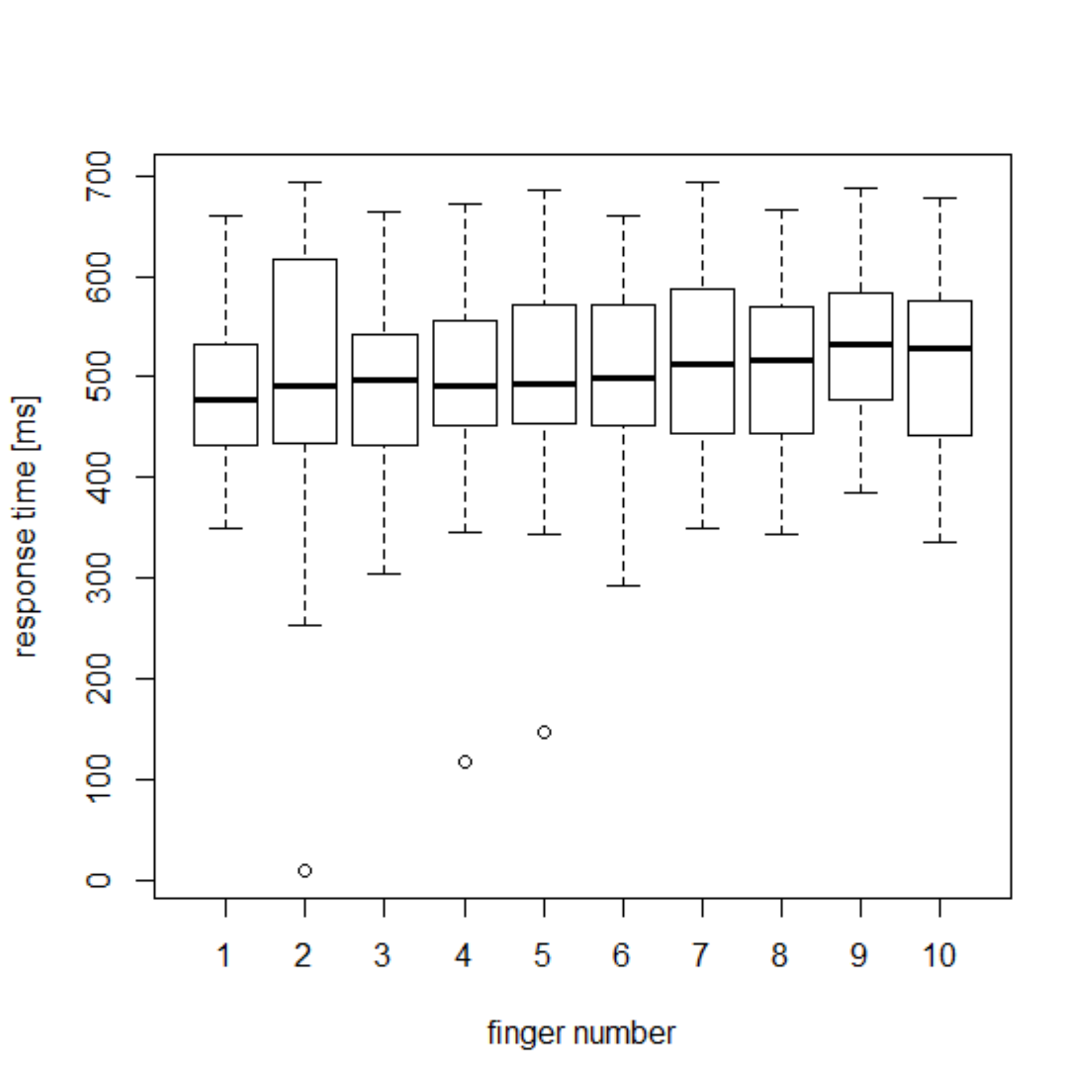}
	\end{center}
	\caption{Results in boxplots (note the overlapping quartiles visualizing no significant differences between medians) of the grand mean averages ($11$ subjects) of the psychophysical experiment response time delays. Each number represents a finger, as depicted in Figure~\ref{fig:10fingers}. The dots represent outliers.}
	\label{fig:allfinger}
\end{figure}

The psychophysical experiment results are summarized in Figure~\ref{fig:allfinger}, where the median response time and the range are depicted for each fingertip as boxplots (see also Figure~\ref{fig:10fingersINTERFACE} for the fingertip numbering). 

The Wilcoxon rank sum tests for pairwise comparisons revealed no differences (at the $0.05$ level) among the median values for all the fingertip pairs of each subject. This result confirms the stimulus similarity since the behavioral responses for all the fingers were basically the same. This finding validates the design of the EEG experiment.

\subsection{EEG Experiment Results}

The results of the EEG experiment are summarized in Table~\ref{tb:bpmr3} and Figure~\ref{fig:feall}. All eleven BCI-naive subjects scored well above the chance level of $10\%$, reaching an ITR in the range from $0.19$~bit/min to $4.46$~bit/min, which may be considered to be a good outcome for experiments with beginners (naive subjects). The ITR was calculated as follows~\cite{bciSPATIALaudio2010}:
\begin{eqnarray}
	&ITR& = V \cdot R\\
	&R& = log_2 N + P\cdot log_2 P + (1-P)\cdot log_2\left(\frac{1-P}{N-1}\right),
\end{eqnarray}
where $R$ stands for the number of bits/selection; $N$ is the number of classes ($10$ in this study); $P$ is the classifier accuracy (see Table~\ref{tb:bpmr3}); and $V$ is the classification speed in selections/minute ($3$~selections/minute for this study). The maximum ITR it was possible for the BCI-naive subjects to achieve in the settings presented was $9.96$~bit/min.

\section{Conclusions}

This paper reports results obtained with a novel ten-command tBCI prototype developed and used in experiments with eleven BCI-naive subjects.  The proposed interface could be used for real-time operation of robotic vehicles. 
The experiment results obtained in this study confirmed the general validity of the tBCI for interactive applications.

In the psychophysical experiment, it is shown that all the tested fingertip zones are equally sensitive to the stimuli and all can be used with the tBCI prototype. The EEG experiment with the prototype has confirmed that tactile stimuli can be used to operate robotic devices with up to ten commands and with the command interfacing rate ranging from $0.19$~bit/min to $4.46$~bit/min for untrained users. 

The results presented offer a step forward in the development of neurotechnology applications. Due to the still not very practical interfacing rate achieved, allowing for only about three commands per minute, the current prototype would  obviously need improvements and modifications. These needs determine the major lines of study for future research. However, even in its current form, the proposed tBCI can be regarded as a practical solution for totally-locked-in (TLS) patients, who cannot use vision or auditory based interfaces due to sensory or other disabilities. 

\section*{Acknowledgments.} 

This research was supported in part by the Strategic Information and Communications R\&D Promotion Program no. 121803027 of The Ministry of Internal Affairs and Communication in Japan, and by KAKENHI, the Japan Society for the Promotion of Science, grant no. 12010738. We also acknowledge the technical support of YAMAHA Sound \& IT Development Division in Hamamatsu, Japan.

\newpage
\section*{Figure Legends}

\begin{description}
	\item[Figure~\ref{fig:10fingers}] The experimental set-up. Tactile stimuli are delivered to the fingertips by ten \textsf{HIAX25C10-8} exciters. The set-up was used in both the psychophysical and the EEG experiments.
	\item[Figure~\ref{fig:10fingersINTERFACE}] The visual instruction screen presented to the subjects during the psychophysical experiment.  Each fingertip was assigned a number encoding a command in the interactive application. The controls on the right side were used by the subject to adjust the stimulus intensity (with a fader), and also by the experimenter to start the experiment (the button with the speaker launches the  \emph{digital-to-analog} signal converter \textsf{MOTU UltraLite-mk3 Hybrid}) and to save/clear the results.
	\item[Figure~\ref{fig:feall}] Grand mean averaged results of the fingertip stimulation EEG experiment for all $11$ subjects. The top panel presents the head topographic plot of the $TARGET$ versus $non-TARGET$ area under the curve (AUC), a measure commonly used in machine learning intra-class discriminative analysis. ($AUC > 0.5$ is usually assumed to be confirmation of feature separability~\cite{book:pattRECO}). The top panel presents the largest difference as obtained from the data displayed in the bottom panel. The topographic plot also depicts the electrode positions. The fact that all the electrodes received similar AUC values (red) supports the initial electrode placement. The second panel from the top presents averaged SEP responses to the $TARGET$ stimuli (note the clear P300 response in the range of $450-700$~ms). The third panel presents averaged SEP responses to the $non-TARGET$ stimuli (no P300 observed). Finally, the bottom panel presents the AUC of $TARGET$ versus $non-TARGET$ responses (again, P300 could easily be identified).
	\item[Figure~\ref{fig:allfinger}] Results in boxplots (note the overlapping quartiles visualizing no significant differences between medians) of the grand mean averages ($11$ subjects) of the psychophysical experiment response time delays. Each number represents a finger, as depicted in Figure~\ref{fig:10fingers}. The dots represent outliers.
\end{description}

\newpage
\section*{Tables}

\begin{table}[H]
	\begin{center}
	\caption{Conditions of the EEG experiment.}\label{tb:p4}
	\begin{tabular}{|l|l|}
	\hline
	Number of subjects					& $11$ \\
	\hline
	Tactile stimulus length			& $100$~ms \\
	\hline
	Stimulus frequency 		& $200$~Hz \\ 
	\hline
	Inter-stimulus-interval (ISI)		& $400$~ms \\
	\hline
	EEG recording system       			& \textsf{g.SAHARA \& g.MOBIlab+} active dry EEG \\ & electrodes system.\\
	\hline
	Number of the EEG channels			& $8$\\
	\hline
	EEG electrode positions				& \textsf{Cz, CPz, P3, P4, C3, C4, CP5, CP6.}\\
	\hline
	Reference and ground electrodes   	& Behind both of the subject's ears\\
	\hline
	Stimulus generation			& $10$ \textsf{HIAX25C10-8} exciters\\
	\hline
	Number of trials for each subject 	& $5$ \\
	\hline	
	\end{tabular} 
	\end{center}
\end{table}

\newpage
\begin{table}[H]
	\begin{center}
	\caption{Interactive application commands encoded with the finger numbers (see also Figure~\ref{fig:10fingers}).}\label{tab:commands}
	\begin{tabular}{c|l}
	~~Finger number~~ 	& ~~Command~~ \\
	\hline
	$1$				& ~~low speed \\
	$2$				& ~~medium speed \\
	$3$				& ~~high speed \\
	$4$				& ~~stop \\
	$5$				& ~~slow speed reverse \\
	$6$				& ~~go left $(-90^\circ)$ \\
	$7$				& ~~go straight--left $(-45^\circ)$ \\
	$8$				& ~~go straight $(0^\circ)$ \\
	$9$				& ~~go straight--right $(45^\circ)$ \\
	$10$				& ~~go left $(90^\circ)$ \\
	\end{tabular} 
	\end{center}
\end{table}

\newpage
\begin{table}[H]
	\begin{center}
	\caption{The fingertip stimulation EEG experiment accuracy and ITR scores. The theoretical chance level was $10\%$. For the classifier, features were derived from the averages of the five ERPs of all the subjects.}\label{tb:bpmr3}
	\begin{tabular}{| c | c | c |}
	\hline
	~Subject number~  	& ~Maximum accuracy~ 	& ITR \\
	\hline  
	$1$					& $20\%$			& ~~$0.19$~bit/min~~ \\
	$2$					& $40\%$			& ~~$1.34$~bit/min~~ \\ 
	$3$					& $50\%$ 		& ~~$2.21$~bit/min~~ \\
	$4$					& $30\%$ 		& ~~$0.66$~bit/min~~ \\
	$5$					& $30\%$ 		& ~~$0.66$~bit/min~~ \\
	$6$					& $40\%$			& ~~$1.34$~bit/min~~ \\
	$7$					& $70\%$ 		& ~~$4.46$~bit/min~~ \\
	$8$					& $40\%$ 		& ~~$1.34$~bit/min~~ \\
	$9$ 					& $20\%$ 		& ~~$0.19$~bit/min~~ \\
	$10$					& $30\%$ 		& ~~$0.66$~bit/min~~ \\
	$11$					& $20\%$			& ~~$0.19$~bit/min~~ \\
	\hline
	\end{tabular} 
	\end{center}
\end{table}

\newpage
\newpage
\section*{Figures}

\begin{thebibliography}{10}

\bibitem{aBCIbirbaumer2010}
Halder, S., Rea, M., Andreoni, R., Nijboer, F., Hammer, E., Kleih, S.,
  Birbaumer, N., K{\"u}bler, A.:
\newblock An auditory oddball brain-computer interface for binary choices.
\newblock Clinical Neurophysiology \textbf{121}(4) (2010)  516--523

\bibitem{maxMSP}
http://cycling74.com/:
\newblock Max 6 (2012)

\bibitem{natureHAPTIC2009}
Johansson, R.S., Flanagan, J.R.:
\newblock Coding and use of tactile signals from the fingertips in object
  manipulation tasks.
\newblock Nature Reviews Neuroscience \textbf{10}(5) (2009)  345--359

\bibitem{Jurcak20071600}
Jurcak, V., Tsuzuki, D., Dan, I.:
\newblock 10/20, 10/10, and 10/5 systems revisited: Their validity as relative
  head-surface-based positioning systems.
\newblock NeuroImage \textbf{34}(4) (2007)  1600--1611

\bibitem{HBCIscis2012hiromuANDtomek}
Mori, H., Matsumoto, Y., Makino, S., Kryssanov, V., Rutkowski, T.M.:
\newblock Vibrotactile stimulus frequency optimization for the haptic {BCI}
  prototype.
\newblock In: Proceedings of The 6th International Conference on Soft Computing
  and Intelligent Systems, and The 13th International Symposium on Advanced
  Intelligent Systems, Kobe, Japan (November 20-24, 2012)  2150--2153

\bibitem{sssrBCI2006}
Muller-Putz, G., Scherer, R., Neuper, C., Pfurtscheller, G.:
\newblock Steady-state somatosensory evoked potentials: suitable brain signals
  for brain-computer interfaces?
\newblock Neural Systems and Rehabilitation Engineering, IEEE Transactions on
  \textbf{14}(1) (March 2006)  30--37

\bibitem{book:eeg}
Niedermeyer, E., {Da Silva}, F.L., eds.:
\newblock Electroencephalography: Basic Principles, Clinical Applications, and
  Related Fields. 5th ed.
\newblock Lippincott Williams \& Wilkins (2004)

\bibitem{p300classifier}
Potes, C.M.:
\newblock {P300 Classifier} (2009)

\bibitem{bci2000book}
Schalk, G., Mellinger, J.:
\newblock A Practical Guide to Brain-Computer Interfacing with {BCI2000}.
\newblock Springer-Verlag London Limited (2010)

\bibitem{bciSPATIALaudio2010}
Schreuder, M., Blankertz, B., Tangermann, M.:
\newblock A new auditory multi-class brain-computer interface paradigm: Spatial
  hearing as an informative cue.
\newblock PLoS ONE \textbf{5}(4) (04 2010)  e9813

\bibitem{book:pattRECO}
Theodoridis, S., Koutroumbas, K.:
\newblock Pattern Recognition. Fourth ed.
\newblock Academic Press (2009)

\bibitem{JNEtactileBCI2012}
van~der Waal, M., Severens, M., Geuze, J., Desain, P.:
\newblock Introducing the tactile speller: an {ERP}-based brain-computer
  interface for communication.
\newblock Journal of Neural Engineering \textbf{9}(4) (2012)  045002

\bibitem{bciBOOKwolpaw}
Wolpaw, J.R., Wolpaw, E.W., eds.:
\newblock Brain-Computer Interfaces: Principles and Practice.
\newblock Oxford University Press (2012)

\end{thebibliography}
\end{document}